\documentclass{article}

\usepackage{PRIMEarxiv}

\usepackage[utf8]{inputenc} 
\usepackage[T1]{fontenc}    
\usepackage{hyperref}       
\usepackage{url}            
\usepackage{booktabs}       
\usepackage{amsfonts}       
\usepackage{nicefrac}       
\usepackage{microtype}      
\usepackage{lipsum}
\usepackage{fancyhdr}       
\usepackage{graphicx}       

\usepackage{multirow}
\usepackage{multicol}
\usepackage{amsmath,graphicx,hyperref}

\title{An Empirical Study for Representations of Videos in Video Question Answering via MLLMs
}

\author{
  Zhi Li, Yanan Wang, Hao Niu, Julio Vizcarra, Masato Taya \\
  KDDI Research Inc. \\
  Fujimino-shi \\
  Japan\\
  \texttt{\{zh-li, wa-yanan, ha-niu, xju-vizcarra, ma-taya\}@kddi.com} \\
}

\begin{document}
\maketitle

\begin{abstract}
Multimodal large language models (MLLMs) have recently achieved remarkable progress in video question answering (VideoQA) by jointly processing visual, textual, and audio information. 
However, it remains unclear which video representations are most effective for MLLMs, and how different modalities balance task accuracy against computational efficiency. 
In this work, we present a comprehensive empirical study of video representation methods for VideoQA with MLLMs. 
We systematically evaluate single-modality inputs—question only, subtitles, visual frames, and audio signals—as well as multimodal combinations, on two widely used benchmarks: VideoMME and LongVideoBench. 
Our results show that visual frames substantially enhance accuracy but impose heavy costs in GPU memory and inference latency, while subtitles provide a lightweight yet effective alternative, particularly for long videos.
These findings highlight clear trade-offs between effectiveness and efficiency and provide practical insights for designing resource-aware MLLM-based VideoQA systems.
\end{abstract}

\keywords{Multimodal large language models \and Video question answering \and Multimodal representations}

\section{Introduction}
\label{sec:intro}
The rapid development of multimodal large language models (MLLMs), such as the GPT series \cite{achiam2023gpt} and the LLaVA-video family \cite{zhang2024llavaov}, has significantly advanced the task of video question answering (VideoQA). 
By jointly processing multiple modalities—including visual frames, audio signals, and textual subtitles—these models achieve strong multimodal reasoning capabilities and surpass traditional deep learning approaches on established benchmarks \cite{yin2024MLLMsurvey1, zhang2024mmsurvey2, BenchSV24Li}. 
As MLLMs continue to scale in both capacity and applicability, understanding how to most effectively represent video content for VideoQA becomes a critical research question.

\noindent
\textbf{Challenge.} 
A variety of methods have been proposed to encode visual, textual, and audio information from videos, yet it remains unclear which modality is most effective for MLLMs in VideoQA tasks \cite{BIMBA25Md,tang2025video,ChenZLLMS25,Zhang24LRVQA}. 
Each modality introduces distinct benefits and limitations: visual frames capture rich spatial and temporal information but impose heavy GPU memory and latency costs; textual subtitles compress semantic content efficiently but may omit fine-grained details; and audio signals provide complementary cues that may or may not align with the query. These differences highlight the importance of carefully selecting and combining modalities in order to maximize both task accuracy and computational efficiency.

\noindent
\textbf{Motivation.} 
Despite the diversity of available methods, a systematic and comprehensive comparison across modalities is still lacking.
Existing works often emphasize a single modality or evaluate representation strategies in isolation, providing limited guidance on how to balance performance gains with computational costs \cite{Liu25UIMLLM,Tang25SFC,Zhang24LRVQA}. 
For example, this work studies the capability of MLLMs on understanding images by analyzing the distributions of attentions between the question and the vision tokens extracted from the image \cite{Liu25UIMLLM}.
Besides, LLoVi introduced a framework to understand long-range video based on the generated captions for the short clips in that video \cite{Zhang24LRVQA}. 
For practitioners aiming to deploy MLLM-based VideoQA systems on long or complex videos, it is crucial to understand the trade-offs between effectiveness and efficiency inherent in different video representations.

To address this gap, this work presents a comprehensive empirical study of video representation methods for VideoQA via MLLMs. 
We investigate the impact of single-modality inputs—such as question only (Q), subtitles (S), and visual frames (V)—as well as multimodal combinations (e.g., Q+S, Q+V, Q+S+V, and configurations including audio). 
We evaluate these settings on two widely used benchmarks, VideoMME \cite{Fu25VideoMME} and LongVideoBench \cite{Wu24LVB}, comparing both their effectiveness in terms of accuracy and their efficiency in terms of GPU VRAM consumption and inference latency. 
Through this analysis, we highlight the relative strengths and weaknesses of each modality, identify practical trade-offs between performance and resource usage, and provide actionable insights for the design of efficient VideoQA pipelines with MLLMs.

\begin{figure*}
\centering
  \centering
  \includegraphics[width=0.90\textwidth]{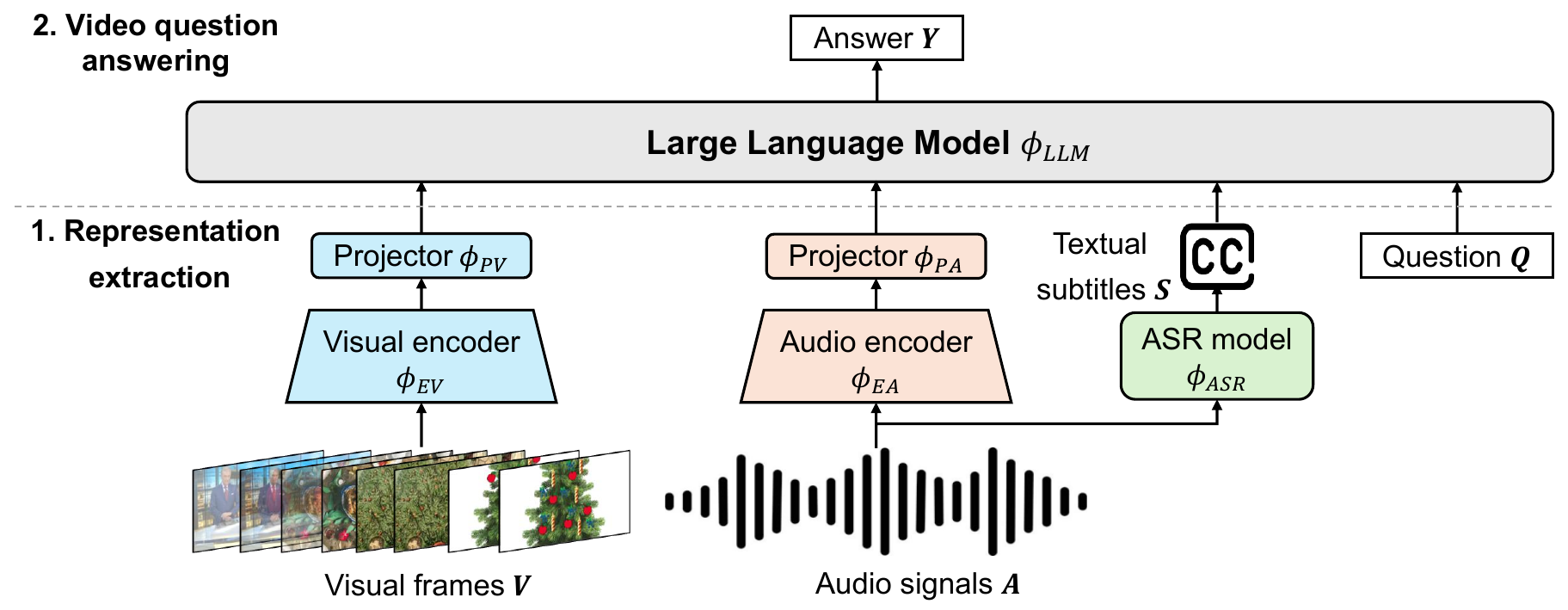}
  \caption{An overview of our framework. It consists of a representation extraction module and a question answering module. The first extracts multimodal video representations and the second inputs those representations into a MLLM to answer questions.}
  \label{fig:framework}
\end{figure*} 

\section{Methodology}
\label{sec:Methodology}
In this section, we elaborate on the details of our framework used to compare the performance of different video representations.
An overview of the framework is shown in Figure \ref{fig:framework}.
As shown in this figure, this framework consists of a representation extraction module and a question answering module. 



\begin{table*}[t]
\centering
\caption{Comparison of different input representations on VideoMME and LongVideoBench, where the input includes questions (Q), subtitles (S), visual frames (V), and audio signals (A). Blod shows the winner. Video-SALMONN 2+ only learns a projector to align audio-text, where the rest parts in this model are equal to Qwen2.5-vl. }
\label{tab:OverallResult}
\resizebox{\textwidth}{!}{
\begin{tabular}{l c | c c c c | c c c c c}
\hline
\textbf{Model} & \textbf{Input} & \multicolumn{4}{c|}{\textbf{VideoMME}} & \multicolumn{5}{c}{\textbf{LongVideoBench}} \\
\cline{3-11}
& & Overall & Short & Medium & Long & Overall & 15s & 60s & 600s & 3600s \\
\hline
Random
& --       & 25.0 & 25.0 & 25.0 & 25.0 & 20.0 & 20.0 & 20.0 & 20.0 & 20.0 \\
\hline
\multirow{4}{*}{Llava-video-7B} 
& Q       & 42.6 & 43.3 & 43.4 & 40.9 & 42.3 & 41.3 & 48.8 & 42.5 & 40.6 \\
& Q+S     & 60.4 & 55.3 & 61.3 & 64.7 & 46.4 & 43.4 & 52.9 & 48.8 & 43.8 \\
& Q+V     & 64.2 & 77.0 & 63.1 & 52.4 & 58.3 & 66.7 & 67.4 & 59.0 & 52.1 \\
& Q+S+V   & \textbf{72.1} & \textbf{80.3} & 69.0 & \textbf{67.0} & 59.3 & 64.6 & 69.2 & \textbf{60.4} & 53.7 \\
\hline
\multirow{4}{*}{Qwen2.5-vl-7B} 
& Q       & 41.1 & 40.3 & 42.6 & 40.3 & 39.3 & 41.8 & 41.3 & 40.3 & 37.1 \\
& Q+S     & 58.0 & 49.6 & 59.0 & 65.4 & 45.8 & 47.1 & 48.8 & 48.8 & 42.2 \\
& Q+V     & 61.7 & 71.6 & 61.6 & 52.0 & 57.0 & 66.7 & 70.3 & 57.5 & 49.3 \\
& Q+S+V   & 70.1 & 75.7 & 68.6 & 66.0 & \textbf{61.4} & \textbf{71.4} & \textbf{70.9} & 60.2 & \textbf{56.0} \\
\hline
Video-SALMONN 2 7B
& Q+A+V   & 67.4 & 79.8 & 65.0 & 57.3 & --   & --   & --   & --   & --   \\
\hline
Video-SALMONN 2+ 7B
& Q+A+V   & 71.1 & 79.0 & \textbf{72.1} & 62.3 & --   & --   & --   & --   & --   \\
\hline
\hline
\multirow{4}{*}{Llava-video-72B} 
& Q       & 48.9 & 47.3 & 50.6 & 48.7 & 47.4 & 50.8 & 54.1 & 47.6 & 44.1 \\
& Q+S     & 66.9 & 59.7 & 68.0 & 72.9 & 51.4 & 48.1 & 56.4 & 54.6 & 48.6 \\
& Q+V     & 70.6 & 81.4 & 68.9 & 61.5 & 61.3 & 71.4 & 74.4 & 60.2 & 54.6 \\
& Q+S+V   & 78.3 & 82.9 & 76.8 & \textbf{75.3} & 62.9 & 74.1 & 75.0 & 62.1 & 56.0 \\
\hline
\multirow{4}{*}{Qwen2.5-vl-72B} 
& Q       & 46.9 & 45.7 & 47.6 & 47.4 & 44.8 & 49.7 & 47.1 & 45.1 & 42.4 \\
& Q+S     & 63.1 & 53.9 & 63.9 & 71.7 & 47.2 & 49.2 & 51.2 & 51.5 & 42.2 \\
& Q+V     & 71.2 & 80.1 & 71.3 & 62.2 & 63.7 & 75.1 & 72.7 & 64.6 & 56.4 \\
& Q+S+V   & \textbf{78.4} & 83.3 & 77.2 & 74.8 & \textbf{65.2} & \textbf{77.2} & 76.2 & \textbf{65.5} & \textbf{57.6} \\
\hline
Video-SALMONN 2+ 72B
& Q+A+V   & 78.3 & \textbf{84.3} & \textbf{79.4} & 71.2 & --   & --   & --   & --   & --   \\
\hline
\end{tabular}
}
\end{table*}

\subsection{Representation Extraction}

To better understand video content, this module extracts variance representations and learns knowledge from multimodal features of videos, including visual frames, audio signals, and textual subtitles.

\vspace{1.0mm}
\noindent
\underline{\textbf{Visual Frames.}}
Due to the limitation of GPU ARAM and the severe latency during question answering process, most MLLMs can process only a limited number of video frames \cite{zhang2024llavaov,BIMBA25Md}.
Given a long video (i.e., a sequence of image frames $\mathbf{I} \in \mathbb{R}^{h \times w \times n}$, most studies uniformly sample a subsequence of frames $\mathbf{V} \in \mathbb{R}^{h \times w \times l}$ from $\mathbf{I}$, 
where $h$ ($w$) is the height (weight) pixels for a single frame and $n$ and $l$ are the number of frames in $\mathbf{I}$ and $\mathbf{V}$, respectively.

After that, the sampled sequence $\mathbf{V}$ is input into a pre-trained visual encoder $\phi_{EV}$ followed by a pre-trained MLP projector $\phi_{PV}$ to obtain visual representations $\mathbf{E_V}$, which is 
\begin{equation}
\mathbf{E_V} = \phi_{PV}(\phi_{EV}(\mathbf{V})).
\end{equation}

\vspace{1.0mm}
\noindent
\underline{\textbf{Audio Signals.}}
Audio is typically paired with video and provides vital complementary information. 
Audio-visual large language models, e.g., GPU-4o \cite{achiam2023gpt} and video-SALMONN-2 \cite{tang2025video}, introduce an additional audio encoder and an audio-text MLP projector to capture audio information and learn audio representations.
Let the raw audio signal of a long video be represented as $\mathbf{A}$.
The signal $\mathbf{A}$ is then input into the audio encoder $\phi_{EA}$ and the MLP projector $\phi_{PA}$, which is
\begin{equation}
\mathbf{E_A} = \phi_{PA}(\phi_{EA}(\mathbf{A})),
\end{equation}
where $\mathbf{E_A}$ denotes audio representations.

\vspace{1.0mm}
\noindent
\underline{\textbf{Textual Subtitles.}} 
Textual captions for a long video can provide necessary
information for MLLMs to correctly understand video content.
However, it is extremely time-consuming and expensive to 
write accurate dense captions manually as well as generating such captions automatically via MLLMs. 

Automatic speech recognition (ASR) provides an alterative choice that recognizes speech in video and transfers it to textual subtitles \cite{ChenZLLMS25,tang2025video}. 
To extract subtitles for long videos, we design a two-stage pipeline that integrates voice activity detector (VAD) for temporal segmentation and ASR for transcription. 
This process ensures robustness against noisy input and maintains synchronization between audio segments and the generated textual subtitles.

Given the raw audio signal of a long video $\mathbf{A} = \{a(t)| t \in [0, T]\}$, 
We apply the WebRTC-based VAD \cite{pywebrtcvad}, which assigns a binary label $y(t) \in \{0, 1\}$, indicating whether speech is present
($y(t)=1$) or absent ($y(t)=0$) at time $t$.
Formally, the speech region set is defined as
\begin{equation}
\mathcal{T} = \{ \,[t_s, t_e] \mid y(t) = 1, \ \forall t \in [t_s, t_e] \,\}.
\end{equation}
This produces a sequence of $N$ short audio segments:
\begin{equation}
\mathcal{C} = \{ c_1, c_2, \ldots, c_N \},
\end{equation}
where each clip $c_i = x(t), \ t \in [t_s^i, t_e^i]$, and $[t_s^i, t_e^i]$ denotes the temporal boundaries of detected speech.

The ASR model, i.e., Whisper-Large-v3 \cite{Radford23Whisper}, is widely used to generate subtitles and extract audio latent representations. 
We, hence, employ Whisper-Large-v3 $\phi_{ASR}$ to generate textual subtitles for each segment $c_i$, which can be formulated as
\begin{equation}
s_i = \phi_{ASR}(c_i).
\end{equation}
Thus, the entire subtitle set for the video is constructed as 
$\mathbf{S} = \{ s_1, s_2, \ldots, s_N \}$.
This pipeline enables efficient subtitle extraction from long videos while maintaining high transcription accuracy and precise temporal alignment.

\subsection{Video Question Answering}
After preparing the video representations, including visual embeddings $\mathbf{E}_I$, audio embeddings $\mathbf{E}_A$, and textual subtitles $\mathbf{S}$, we feed each representation (or their combination) together with the given question $\mathbf{Q}$ into a pre-trained frozen LLM to generate the final textual answer.  
Formally, this process can be expressed as:
\begin{equation}
    \mathbf{Y} = \phi_{\text{LLM}}\bigl( \mathcal{P}, \phi_{ET}(\mathbf{Q}) \bigr),
\end{equation}
where $\mathcal{P} \in \bigl\{ \mathbf{E}_I, \, \mathbf{E}_A, \, \phi_{ET}(\mathbf{S}) \bigr\}$ denotes a single modality or a multimodal combination of video representations,  
$\mathbf{Q}$ is the input question,  
$\phi_{ET}$ is the text encoder, 
and $\mathbf{Y}$ is the generated textual answer.



\section{Experiments}
\label{sec:Exp}

\subsection{Experimental Setup}
\vspace{1.0mm}
\noindent
\textbf{Datasets.}
We validate the performance of a single modality or a multimodal combination of video representations on two widely used long \textit{VideoQA} benchmarks, i.e., VideoMME \cite{Fu25VideoMME} and LongVideoBench (Validation set) \cite{Wu24LVB}. 

VideoMME contains $900$ YouTube videos, including $300$ short videos ($<$2 minutes), $300$ medium videos ($4-15$ minutes), and 300 long videos ($30-60$ minutes).
It covers $6$ key domains and includes $3,000$ four-choice question tasks ranging from easy object recognition to complex temporal reasoning.

LongVideoBench (Val.) contains $753$ web videos, categorized into 4 groups based on the duration (seconds) of the video: $15s$, $60s$, $600s$, and $3600s$.
It covers 4 themes and includes $1,337$ five-choice questions ranging from easy object recognition to complex multiple scene reasoning.

\vspace{1.0mm}
\noindent
\textbf{MLLM Backbone.}
We evaluated several open MLLMs that are available for commercial use and achieve state-of-the-art performance on most VQA benchmarks, including Llava-video-7B \cite{zhang2024llavaov}, Llava-video-72B \cite{zhang2024llavaov}, Qwen2.5-VL-7B \cite{Bai25qwen25}, Qwen2.5-VL-72B \cite{Bai25qwen25}, Video-SALMONN 2 7B \cite{tang2025video}, Video-SALMONN 2+ 7B \cite{tang2025video}, and Video-SALMONN 2+ 72B \cite{tang2025video}. 

Both Llava-video and Qwen2.5-VL pre-train a visual encoder and a text encoder that can process visual frames and textual subtitles, while
Video-SALMONN 2 and Video-SALMONN 2+ pre-train an additional audio encoder that can process raw audio signal.

\vspace{1.0mm}
\noindent
\textbf{Implementation.} We employ the lmms-eval \cite{Zhang25LMMeval} framework to conduct systematic evaluations of the LLaVA-video and Qwen2.5-VL models. 
All hyperparameters are kept at their default values in order to ensure fair and reproducible comparisons across different model backbones and input modalities. 
For the Video-SALMONN series of models, we directly adopt the results reported by the original authors \cite{tang2025video}.

\begin{figure}[t]
\centering
  \centering
  \includegraphics[width=0.45\textwidth]{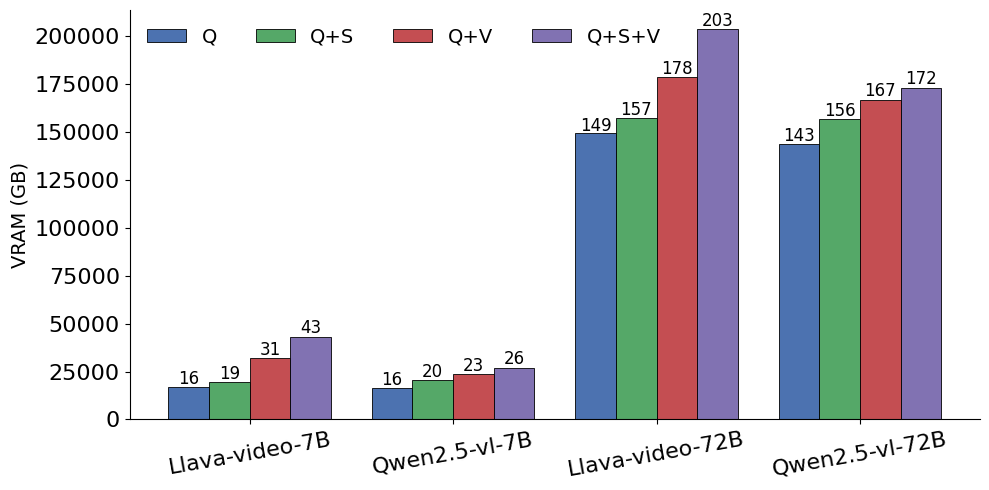}
  \caption{GPU memory consumption across different input modalities on VideoMME.}
  \label{fig:memory}
\end{figure}

\subsection{Experimental Result}
\noindent
\textbf{(1) How does different modalities of video representations affect the accuracy of VideoQA?}


Table \ref{tab:OverallResult} reports the comparison results of inputting single modality or different combinations of modalities on VideoMME and LongVideoBench.
From this table, we have the following observations. 

All 7B (72B) MLLMs achieve at least $15\%$ ($20\%$) improvements ,compared to random selection, even prompting only questions and providing no video information.
These improvements are gained by the commonsense knowledge learned during the training process of MLLMs, where similar phenomenons are also reported in recent studies \cite{Liu25UIMLLM,Tang25SFC}. 

In VideoMME, subtitles are more effective than visual frames and audio signals in boosting the accuracy of VideoQA for long videos.
This is because MLLMs can process a small number of frames due to the limitation of token length. 
Also, MLLMs, as a language model, naturally understand textual subtitles better compared to audio signal.

The combination of subtitles and visual frames achieves the best performance in most cases, especially outperforming the combination of audio signal and visual frames over $3.7\%$ and $3.6\%$ based on Qwen-2.5-vl series models.
This finding provides an alterative choice to boost performance on VideoQA for MLLMs that can not process audio signals.

\noindent
\textbf{(2) How does different modalities of video representations consume GPU ARAM?}

We further examine GPU memory consumption across different input modalities and report the results in Figure \ref{fig:memory}. 
The results reveal that textual subtitles introduce only a modest increase in VRAM usage compared to using questions alone, reflecting their relatively lightweight sequence length. 
In contrast, visual frames substantially dominate memory demand, nearly doubling the usage for 7B-scale models and adding over $20\%$ overhead for 72B-scale models. 
When subtitles are combined with visual inputs, the memory footprint reaches its maximum, though the additional cost from subtitles is minor relative to the visual component. 
These findings indicate that visual representations are the primary factor in GPU VRAM consumption, while textual subtitles contribute limited overhead.

\noindent
\textbf{(3) How does different modalities of video representations impact inference latency?}

\begin{figure}[t]
\centering
  \centering
  \includegraphics[width=0.45\textwidth]{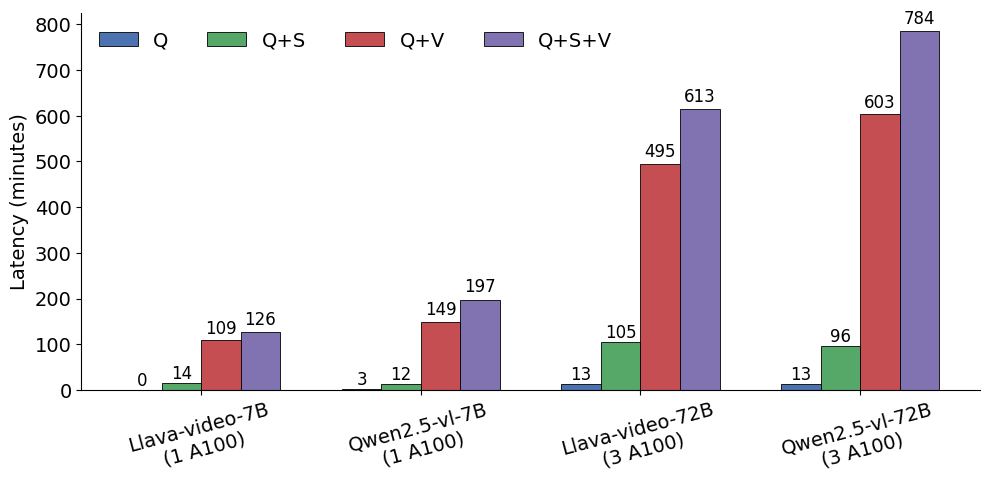}
  \caption{Inference latency across different input modalities on VideoMME. We inference 7B and 72B models on 1 and 3 NVIDIA A100 GPUs, respectively}
  \label{fig:latency}
\end{figure}

We analyze inference latency across different input modalities and report the results in Figure \ref{fig:latency}.
Using only question tokens achieves the lowest latency, as expected, with runtimes of just a few seconds to minutes across model scales. 
Incorporating textual subtitles introduces a moderate increase, reflecting the additional sequence length but remaining relatively efficient compared to visual processing. 
By contrast, the inclusion of visual frames substantially dominates inference latency, pushing runtimes into hours for both 7B and 72B models. 
When subtitles are combined with visual inputs, latency reaches its maximum, though the incremental overhead from subtitles is modest relative to the dominant visual cost. 
These findings indicate that visual representations are the primary bottleneck in inference latency, whereas subtitles add only limited additional overhead.

\section{Conclusion}
\label{sec:typestyle}
In this paper, we presented a empirical study about how different representations impact both the effectiveness and efficiency of MLLMs on VideoQA tasks. 
Our results show that visual frames substantially improve accuracy but dominate GPU VRAM usage and inference latency, while textual subtitles provide a lightweight yet informative alternative that is especially effective for long videos. Moreover, combining subtitles with visual frames consistently yields the strongest performance across benchmarks, though at the expense of additional computational cost.
These findings suggest that the choice of modality should be guided not only by task requirements but also by the available computational budget.

\bibliographystyle{unsrt}  
\bibliography{templateArxiv}

\end{document}